\documentclass[12pt]{article}
\usepackage{graphicx}
\usepackage{placeins}

\newcommand\pubnumber{SNSN-323-63}
\newcommand\pubdate{\today}

\def\institute{Deutsches Elektronen-Synchrotron (DESY)\\
Notkestra{\ss}e 85, 22607 Hamburg, Germany}

\def\Title#1{\begin{center} {\Large #1 } \end{center}}
\def\Author#1{\begin{center}{ \sc #1} \end{center}}
\def\Address#1{\begin{center}{ \it #1} \end{center}}

\newcommand\pubblock{\rightline{\begin{tabular}{l} \pubnumber\\
         \pubdate  \end{tabular}}}
\newenvironment{Abstract}{\begin{quotation}  }{\end{quotation}}
\newenvironment{Presented}{\begin{quotation} \begin{center} 
             PRESENTED AT\end{center}\bigskip 
      \begin{center}\begin{large}}{\end{large}\end{center} \end{quotation}}

\def\beq{\begin{equation}}
\def\eeq#1{\label{#1}\end{equation}}
\def\eeqn{\end{equation}}

\def\beqa{\begin{eqnarray}}
\def\eeqa#1{\label{#1}\end{eqnarray}}
\def\eeqan{\end{eqnarray}}

\let\bar=\overbar

\def\Dslash{\not{\hbox{\kern-4pt $D$}}}
\def\dslash{\not{\hbox{\kern-2pt $\del$}}}

\def\msb{{\bar{\ssstyle M \kern -1pt S}}}

\begin{document}
\begin{titlepage}
\pubblock

\vfill
\Title{CMS measurements of EFT parameters with top quarks} 
\vfill
\Author{Nicolas Tonon on behalf of the CMS Collaboration}
\Address{\institute}
\vfill
\begin{Abstract}
Measurements carried out or reinterpreted within the framework of effective field theory (EFT) will constitute a key component of the LHC legacy in the quest towards physics beyond the standard model. 
Numerous EFT measurements were already published by the CMS Collaboration.
In the top quark sector, four distinct approaches to EFT have been identified, each successfully applied in several analyses.
We review the pros and cons of each approach, while illustrating them with recent analyses in which they were adopted.
\end{Abstract}
\vfill
\begin{Presented}
$13^\mathrm{th}$ International Workshop on Top Quark Physics\\
Durham, UK (videoconference), 14--18 September, 2020
\end{Presented}
\vfill
\end{titlepage}
\def\thefootnote{\fnsymbol{footnote}}
\setcounter{footnote}{0}

\section{Introduction}
The standard model (SM) of particle physics predicts with great accuracy a tremendous number of experimental results 
covering a wide energy range up to the TeV scale accessible at the CERN LHC~\cite{LHC}.
Still, it does not provide explanations for several key observations such as the existence of dark matter and energy, or the masses of neutrinos.
More generally, there exists a number of indications that the SM only corresponds to a low-energy approximation to a more fundamental theory beyond the standard model (BSM). 

Various BSM models postulate the existence of new particles or mechanisms to address these shortcomings.
However the SM does not predict the energy scale at which new physics may appear, 
and extensive research efforts notably carried out by the ATLAS~\cite{ATLAS} and CMS~\cite{CMS} experiments at the LHC covering a large phase space region did not yet indicate the presence of new physics.

The large top quark mass of about 173 GeV and its Yukawa coupling to the Higgs boson close to unity suggest that it may play a special role within the SM, and that its closer study may shed light on the electroweak symmetry breaking mechanism. 
Many BSM theories predict sizeable deviations of the top quark's couplings with respect to SM predictions.
Moreover, most of the canonical top quark processes have now reached the precision era at the LHC, 
and their uncertainties are systematics-dominated.
This motivates carrying out an ambitious research program in the top quark sector in order to reveal new physics effects indirectly through precision measurements.

\section{Effective field theory}

One of the main theoretical frameworks to interpret potential deviations in a model-independent way is that of effective field theory (EFT). 
One key assumption of this approach is that the new physics is characterized by some unknown energy scale $\Lambda$ way beyond the energy reach of the LHC. Under this assumption, the SM Lagrangian is expanded with additional operators of higher mass dimensions ($d > 4$) representing new interactions between SM fields, whose coupling strengths are described by Wilson coefficients (WCs). Constraints placed on WCs may then be mapped onto any UV-complete model.
An EFT thus offers a well-motivated and general framework to maximize the discovery potential of massive BSM states at the LHC and beyond.

For a given basis of operators, i.e. any complete and minimal set of operators at a given order, there is a large number of operators to consider: for instance there are already 59 independent operators impacting the top quark at dimension 6, and this number grows exponentially with mass dimension.
Since constraining this huge parameter space would require a wealth of data and the combination of a large number of analyses targeting different final states (which is the ultimate goal), it is natural to make motivated assumptions regarding the nature of new physics to restrict the considered phase space. 
Most LHC EFT analyses up to now have considered dimension-6 operators only, which are in general expected to describe most new physics effects since higher-order operators are suppressed by powers of $\Lambda$; and operators of dimension 5 or 7 lead to lepton flavor violation and are only relevant in specific analyses.
Most useful guidelines and prescriptions on relevant assumptions are provided in a note arising from the LHCTopWG~\cite{saavedra2018interpreting}, which summarizes the fruitful outcomes of a collaboration between theorists and experimentalists.

\section{Different approaches to EFT}
Within the CMS Top group, four different strategies have been identified and employed so far to constrain EFT operators. 
They can broadly be classified on a spectrum ranging from post-measurement reinterpretations towards direct EFT measurements carried out at the detector level.
In the following, we detail each approach with its pros and cons, and provide recent examples of CMS top quark analyses in which they were adopted.

\subsection{Reinterpretation of an inclusive measurement}
A first approach consists in reinterpreting a cross section measurement a posteriori. 
Its value can be parameterized with the EFT operators of interest, which makes it possible to constrain their WCs in a straightforward manner.
The main advantages of this approach are its good scalability and its ease of combination with other measurements obtained with any experiment. 
It also does not require the generation of any dedicated Monte-Carlo simulated sample including the impact of EFT operators.
On the other hand, such reinterpretations typically rely on assumptions regarding new physics, which usually does not only impact the cross section itself, but also the kinematic distributions of the objects used in an analysis for event selection and signal extraction.

This approach was adopted to reinterpret the cross section measurement of the four-top process by CMS~\cite{4top}. 
This analysis is particularly challenging since the signal has an expected SM cross section $\sigma(SM) = 9.2$~fb, 5 orders of magnitude smaller than that of the main $t\bar{t}$ background. 
The four-top process is highly sensitive to four-heavy-quark operators. 
The signal cross section was parameterized at the generator level with several EFT operators as:
\begin{equation}
 \sigma_{t\bar{t}t\bar{t}} = \sigma_{t\bar{t}t\bar{t}}^{SM} + \frac{1}{\Lambda^2} \sum_k C_k \sigma_k^{(1)} + \frac{1}{\Lambda^4} \sum_{j \leq k} C_j C_k \sigma_{j,k}^{(2)} \, , 
\end{equation}
where the linear terms $C_k \sigma_k^{(1)}$ represent the interference terms of the SM and dimension-6 EFT contributions, while the quadratic terms include two components: the squared contributions from diagrams containing one EFT operator, and the interference terms for two diagrams both including the insertion of one operator.
This parameterization was used to translate the measured cross section into 95\% CL upper limits on each individual WC, while marginalizing over the other operators.

\subsection{Reinterpretation of an unfolded differential measurement}
A second approach consists in reinterpreting a differential measurement unfolded at the generator level.
Such reinterpretations may be combined with other results if bin-to-bin correlation matrices are provided, 
and require the generation of differential Monte-Carlo sample including EFT effects at the generator level. 
Differential measurements of quantities sensitive to EFT make it possible to also exploit shape information and typically lead to tighter constraints compared to inclusive ones, but ignore the effects that new physics may have on the detector acceptance and selection efficiencies.

This procedure was followed in the CMS measurement of the differential $t\bar{t}$ cross section as a function of kinematic observables of the top quarks, of its decay products and of the $t\bar{t}$ system~\cite{tt1}.
This analysis targeted the dileptonic opposite-sign final state. 
The measurements were unfolded both to the parton and particle levels, in full and fiducial phase spaces respectively. 
The angular separation between the two leptons $\Delta \phi (\ell\ell)$ was simulated at generator level using the RIVET framework and parameterized with the $\mathcal{O}_{tG}$ operator at next-to-leading order (NLO) in QCD. 
This operator modifies SM vertices and introduces new coupling structures, thus impacting both the yield and kinematics of the $t\bar{t}$ process. It is directly related to the top quark chromomagnetic dipole moment (CMDM), which is predicted to have a small value within the SM and is modified in several BSM models (2HDM, SUSY, technicolor, etc.).
Limits were set on $\mathcal{O}_{tG}$ at the particle level at NLO. The higher accuracy in QCD was found to enhance the effects of this operator, and to significantly reduce theoretical scale uncertainties compared to LO predictions.

Using the same data and targeting the same final state, a CMS measurement~\cite{tt2} of the differential $t\bar{t}$ cross section performed as a function of polarization and spin correlation observables was used to constrain $\mathcal{O}_{tG}$ via a simultaneous $\chi^2$ fit to several kinematic observables sensitive to spin correlations.
Exploiting these powerful variables allowed to improve the sensitivity to $\mathcal{O}_{tG}$ by about 30\% compared to the previous analysis.

\subsection{Hybrid measurement performed at detector level}
A third approach dubbed ``hybrid measurement'' relies on the EFT parameterization of yields or differential distributions at the generator level; this parameterization is then translated to the detector level under SM assumptions to extract the results. 

This approach was adopted in a search for new physics targeting the $t\bar{t}$ and $tW$ processes in the dileptonic final state~\cite{np}.
Limits on six different operators were extracted from a simultaneous fit to counting experiments and neural network discriminants in several categories. This constituted a first important step towards more global fits wherein EFT effects are considered in more than one process.

A CMS measurement of the $t\bar{t}Z$ cross section in $3\ell$ and $4\ell$ final states set limits on several EFT operators using a more involved procedure~\cite{ttz}: first, LO generator level samples were produced on a fine grid over the theory phase space (i.e. both at SM and non-SM points); ratios were computed at different points with respect to the SM prediction to scale the distributions of interest, before applying any event selection; finally, these weights were translated to the detector level and applied to the nominal NLO signal sample to emulate the EFT contributions. The validity of the entire procedure was verified in closure tests.
Two-dimensional differential distributions were used to extract limits both within the EFT and anomalous coupling frameworks.

\subsection{Direct EFT measurement}
Finally the direct measurement approach minimizes the number of SM assumptions to make, and makes it possible to consider EFT effects in all sensitive processes at once. It offers maximal control over correlations and systematic uncertainties. 
However, it requires the generation of samples including EFT effects up to the detector level.

\begin{figure}[!h]
\centering
\includegraphics[height=9cm]{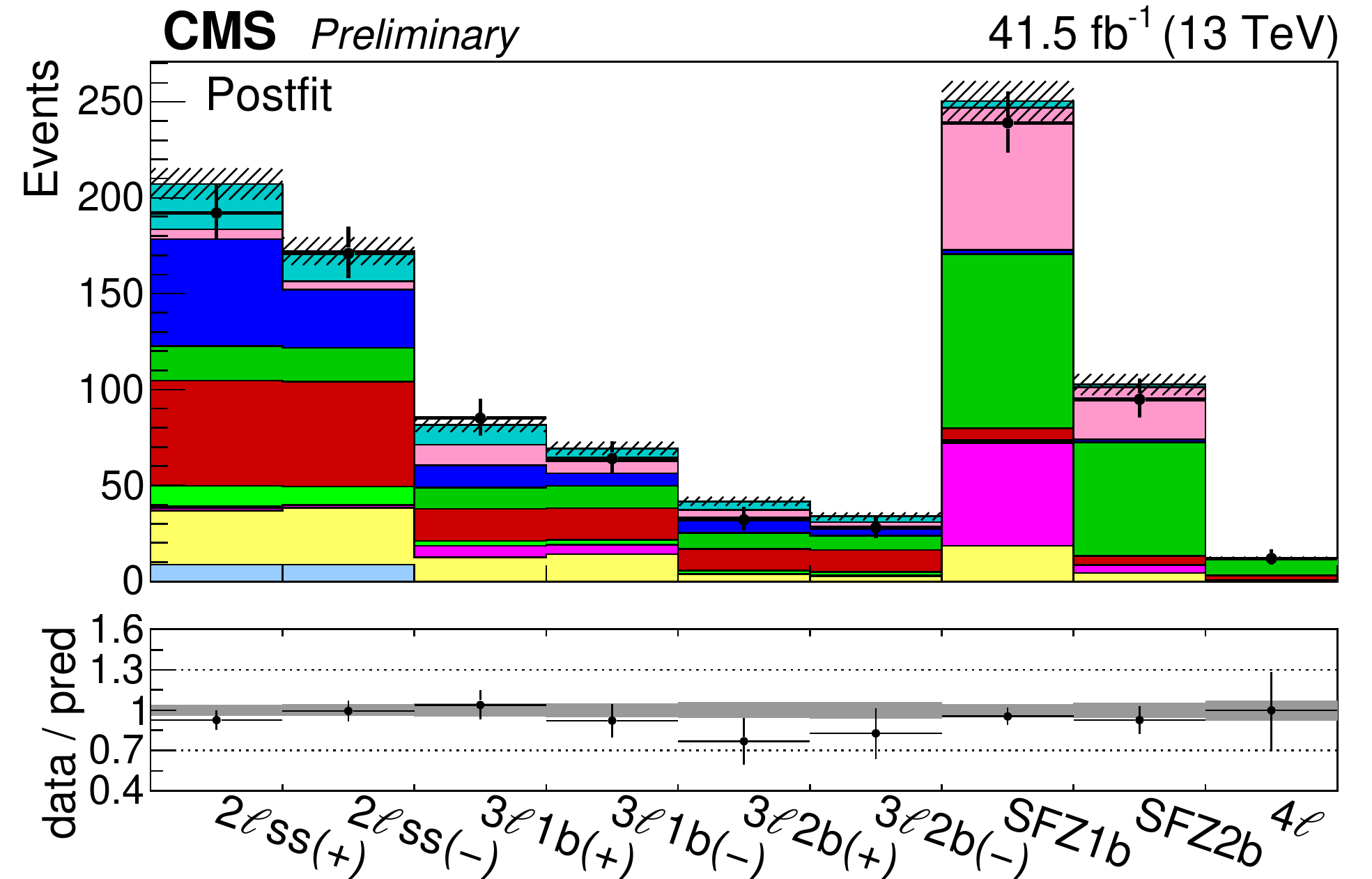}
\caption{Postfit yields in several multilepton categories from Ref.~\cite{top19001}. 
The yields of five top quark processes are parameterized with 16 different WCs, which get constrained in a fit to the data.}
\label{fig:1}
\end{figure}

A recent CMS analysis~\cite{top19001} employed this approach for the first time in the top quark sector to constrain a set of 16 relevant operators. 
It considered EFT effects in five associated production modes of the top quark with gauge and Higgs bosons ($t\bar{t}Z$, $t\bar{t}W$, $tZq$, $t\bar{t}H$, $tHq$) in multilepton final states.
Simulated samples including EFT effects were passed through a full simulation of the CMS detector, and events were categorized based on lepton, jet and b jet multiplicities to enhance the separation between different processes. 
The postfit yields are shown for different categories in Fig.~\ref{fig:1}.
Both 1D and 2D limits were extracted for each operator, while either setting the other operators to zero or profiling them.
This represents an important step towards direct measurements including EFT effects in all relevant processes.

\FloatBarrier 

\begin{thebibliography}{99}

\bibitem{LHC}
L. Evans and P. Bryant, JINST 3 (2008) S08001.

\bibitem{ATLAS}
ATLAS Collaboration, JINST 3 (2008) S08003.

\bibitem{CMS}
CMS Collaboration, JINST 3 (2008) S08004. 


\bibitem{saavedra2018interpreting}
J. A. Aguilar Saavedra et al., CERN-LPCC-2018-01 (2018).

\bibitem{4top}
CMS Collaboration, J. High Energ. Phys. 2019, 82 (2019). 

\bibitem{tt1}
CMS Collaboration, J. High Energ. Phys. 2019, 149 (2019). 

\bibitem{tt2}
CMS Collaboration, Phys. Rev. D 100 , 07 (2019).

\bibitem{np}
CMS Collaboration, Eur. Phys. J. C 79, 886 (2019).

\bibitem{ttz}
CMS Collaboration, J. High Energ. Phys. 2020, 56 (2020). 

\bibitem{top19001}
CMS Collaboration, CMS-PAS-TOP-19-001 (2020). 

\end{thebibliography}
\end{document}